\documentclass[12pt]{article}
\usepackage{graphicx}
\usepackage{lscape}
\usepackage{epsfig}
\usepackage{citesort}
\usepackage{amssymb}
\usepackage{amsmath}
\usepackage{multirow}
\begin{document}
\title{ {\Large \textbf{A Cosmological Exact Solution of Complex Jordan-Brans-Dicke Theory
and its Phenomenological Implications } } }

\author{ {\small M. Ar\i k}$^\dag$, {\small M. \c{C}al\i k}$^\ddag$, {\small N.Kat\i rc\i}$^*$\\
{\small ${}\dag$ Department of Physics, Bogazici University, Bebek
Istanbul,Turkey, } \\
{\small ${}^*$ Physics Division, Faculty of Arts and Sciences,
Do\u gu\c s University}\\
{\small Ac{\i}badem-Kad{\i}k\"oy, 34722 Istanbul, Turkey}}

\begin{titlepage}
\maketitle
\thispagestyle{empty}
\begin{abstract}
When Brans-Dicke Theory is formulated in terms of the Jordan scalar field $%
\varphi$, dark energy is related to the mass of this field. We
investigate the
solution which is relevant for the late universe. We show that if $%
\varphi$ is taken to be a complex scalar field then an exact
solution of the vacuum equations shows that Friedmann equation
possesses a term, proportional to the inverse sixth power of the
scale factor, as well as a constant term. Possible interpretations
and phenomenological implications of this result are discussed.
\end{abstract}
~~~Keyword(s): Jordan-Brans-Dicke Theory, supernovae Type $1A$,
dark energy
\end{titlepage}
\section{Introduction}
Jordan in $1959$ proposed a scalar field $\varphi $ and replaced
the gravitational constant in the Einstein-Hilbert action by a
term proportional to $\varphi ^{2}$ \cite{jordan1,jordan2}. Brans
and Dicke redefined the scalar field so that $\square \phi =0$ as
a vacuum solution \cite{brans}. Thus, in Jordan-Brans-Dicke
theories (JBD), the counterpart of the
gravitational coupling term, $1/16\pi$$G_{N}$, is replaced by $\phi $ or $%
\varphi ^{2}/8\omega $, which may be a function of space and/or
time. For an isotropic homogeneous cosmology, which evolves in
time, scalar field is a function of only time. JBD gravity has
been used extensively to develop dark energy models
\cite{bd1,bd2,bd3,bd4,bd5,bd6} which usually involve a scalar
potential which is adjusted to fit observed cosmology. In
contrast; we will use a simple model where the JBD field is
complex and besides the kinetic term, it contains only a standard
mass term for the Jordan field $\varphi $. This is in contrast to
most cosmological models based on Jordan-Brans-Dicke theory where
an arbitrary potential $V(\varphi)$ is assumed and then the
potential is adjusted to give the desired solution. One possible
motivation for considering a phase for the $JBD$ field is that it
could affect the strong $QCD$ parity violating phase $\chi$
\cite{axion1,axion2} or the CP violating phase of the quark mass
matrix \cite{cabibbo,kobayashi} or neutrino mixing
\cite{neutrinomixing}. These may be related to dark matter. While
in Section II, an exact cosmological solution is displayed and its
stability analysis is made for the vacuum case, in section III, it
is interpreted as the effect of the proposed complex scalar field
to the evolution of the late time universe. We also give a brief
discussion of possible relationship of the phase of the complex
scalar field. We will show that the dark energy component, in
addition to the constant term makes a contribution to the
Friedmann equation, which is proportional to the inverse sixth
power of the scale factor.

We use a metric signature $(+---)$ and the Jordan formalism. The
lagrangian densities of the Jordan and Brans-Dicke language are
related by,
\begin{eqnarray}
\pounds _{BD}=-{\phi }R+\frac{\omega }{\phi }g^{\mu \nu }\partial
_{\mu }\phi \partial _{\nu }\phi  \nonumber \\
 =-\frac{\varphi
^{2}}{8\omega }R+\frac{1}{2}g^{\mu \nu }\partial _{\mu }\varphi
\partial _{\nu }\varphi =\pounds _{JBD}.
\label{lagbd}
\end{eqnarray}%
We add a mass term and a phase $\chi $,
\begin{equation}
\varphi =\varphi _{1}+i\varphi _{2}=\left\vert \varphi \right\vert
e^{i\chi }
\end{equation}%
so that the lagrangian density with a complex massive scalar field
becomes,
\begin{equation}
\pounds =-\frac{\varphi \varphi ^{\ast }}{8\omega
}R+\frac{1}{2}g^{\mu \nu
}\partial _{\mu }\varphi \partial _{\nu }\varphi ^{\ast }-\frac{1}{2}%
m^{2}\varphi \varphi ^{\ast }  \label{esas}
\end{equation}%
which can also be expressed as
\begin{eqnarray}
\pounds =-\frac{\left\vert \varphi \right\vert ^{2}}{8\omega
}\left[ R-4\omega g^{\mu \nu }\partial _{\mu }\chi \partial _{\nu
}\chi +4\omega m^{2}\right] \nonumber \\
+\frac{1}{2}g^{\mu \nu }\partial _{\mu }\left\vert \varphi
\right\vert \partial _{\nu }\left\vert \varphi \right\vert .
\label{cjbd}
\end{eqnarray}

\bigskip The action is defined by
\begin{equation}
S=\int d^{4}x\sqrt{-g}\pounds +S_{M}.  \label{action}
\end{equation}%
When this action is varied with respect to the metric and the
complex scalar field, the equations of motion, for a
Friedmann-Robertson-Walker metric and perfect fluid
energy-momentum tensor $T_{\nu }^{\mu }=diag\left( \rho
,-p,-p,-p\right) ,$ reduce to the following:
\begin{eqnarray}
\frac{3}{4\omega }\,{\varphi \varphi }^{\ast }\,\left( \frac{\dot{a}^{2}}{%
a^{2}}+\frac{k}{a^{2}}\right)
-\frac{1}{2}\,\dot{\varphi}\dot{\varphi}^{\ast
}-\frac{1}{2}\,m^{2}\,\varphi {\varphi }^{\ast }\nonumber \\
+\frac{3}{4\omega }\,\frac{%
\dot{a}}{a}\,(\dot{\varphi}\,{\varphi }^{\ast }+\varphi {\dot{\varphi}}%
^{\ast })=\rho _{M}  \label{rho}
\end{eqnarray}%
\begin{eqnarray}
\frac{-1}{4\omega }{\varphi \varphi }^{\ast }\left( 2\frac{\ddot{a}}{a}+%
\frac{\dot{a}^{2}}{a^{2}}+\frac{k}{a^{2}}\right)-\frac{1}{2\omega}\,\frac{\dot{a}}{a}\,\,(\dot{\varphi}\,{\varphi
}^{\ast }+\varphi {\dot{\varphi}}^{\ast })\nonumber \\
 -\frac{1}{4\omega }(\ddot{\varphi}\,{\varphi
}^{\ast} +\varphi {\ddot{\varphi}}^{\ast })\,-\left(
\frac{1}{2}+\frac{1}{2\omega }\right) \,{
\dot{\varphi}\dot{\varphi}}^{\ast }+\frac{1}{2}\,m^{2}\,\varphi {\varphi }%
^{\ast }=p_{M}  \label{p}
\end{eqnarray}%
\begin{eqnarray}
\ddot{\varphi}+3\,\frac{\dot{a}}{a}\,\dot{\varphi}+\left[ m^{2}-\frac{3}{%
2\omega }\left( \frac{\ddot{a}}{a}+\frac{\dot{a}^{2}}{a^{2}}+\frac{k}{a^{2}}%
\right) \right] \,\varphi =0. \label{fi}
\end{eqnarray}%
Equation \ref{fi}, being complex, is equivalent to two real
equations. H and $F_{1}+iF_{2}$ are respectively defined as the
fractional rate of changes
 of the scale size and the JBD scalar field $\varphi $.
\begin{equation}
H=\frac{\dot{a}}{a}\,\,\,\,\,,\,\,\,F_{1}+iF_{2}=\frac{\dot{\varphi}}{%
\varphi }  \label{H}
\end{equation}%
where $F_{1}$ and $F_{2}$ are real, and defined as,
\begin{eqnarray}
F_{1}=\left\vert \varphi \right\vert ^{-1}\frac{d\left\vert
\varphi \right\vert}{dt}=\frac{\dot{\varphi _{1}}\varphi
_{1}+\dot{\varphi _{2}}\varphi
_{2}}{\varphi _{1}^{2}+\varphi _{2}^{2}}=-\frac{1}{2}\frac{\dot{G_{N}}}{G_{N}%
}  \label{f1}
\end{eqnarray}%
\begin{equation}
F_{2}=\frac{\dot{\varphi _{2}}\varphi _{1}-\dot{\varphi _{1}}\varphi _{2}}{%
\varphi _{1}^{2}+\varphi _{2}^{2}}=\dot{\chi}.  \label{f2}
\end{equation}%
For spatially flat (k=0) universe, we will obtain solutions which give H, F$%
_{1}$, $F_{2}$ as a function of the scale size $a$. After
rewriting equations \ref{rho}, \ref{p}, and \ref{fi} in terms of
H, $F_{1}$, $F_{2}$, and their derivatives with respect to the
scale size of the universe, a, we get the following equations:
\begin{equation}
3H^{2}-2\omega (F_{1}^{2}+F_{2}^{2})+6HF_{1}-2\omega
m^{2}=\frac{4\omega \rho _{M}}{\left\vert \varphi \right\vert
^{2}}  \label{density}
\end{equation}%
\begin{eqnarray}
3H^{2}+(2\omega +4)F_{1}^{2}+2\omega F_{2}^{2}+4HF_{1}
+2aH(F_{1}^{\prime }+H^{\prime })-2\omega m^{2}=\frac{-4\omega
p_{M}}{\left\vert \varphi \right\vert ^{2}}  \label{pressure}
\end{eqnarray}%
\begin{eqnarray}
-6H^{2}+2\omega F_{1}^{2}-2\omega F_{2}^{2}+6\omega HF_{1}
+2a\omega HF_{1}^{\prime } -3aHH^{\prime }+2\omega m^{2}=0
\label{fi1}
\end{eqnarray}%
\begin{equation}
(4\omega F_{1}+6\omega H)F_{2}+2\omega aHF_{2}^{\prime }=0
\label{fi2}
\end{equation}%
where prime denotes derivative with respect to $a$. \bigskip

\section{The exact cosmological solutions for dust dominated  and the vacuum case }

Exact cosmological solutions of JBD theory can be obtained by
analyzing symmetries of the field equations \cite{sheftel}.
Perturbative solutions were found in \cite{cifter,calik}. We
propose an ansatz for finding the exact solutions for the late
time universe. We will show that an exact solution of the
cosmological equations can be obtained by an ansatz for the matter
dominated case and vacuum case.
\begin{equation}
F_{1}(a)=\frac{H(a)}{2(\omega +1)}  \label{f1sol}
\end{equation}%
\begin{equation}
\frac{|\varphi| ^{\prime }}{|\varphi| }=\frac{F_{1}}{aH}
\label{fidotfi}
\end{equation}%
By integrating equation %
\ref{fidotfi} with the ansatz yields
\begin{equation}
|\varphi ^{2}|=|\varphi _{0}^{2}|\left[ \frac{a}{a_{0}}\right] ^{\frac{1}{%
(1+\omega )}}.  \label{fikare}
\end{equation}%
Since $\omega >10^{4}$ \cite{timedel0,timedel1,timedel2,timedel3}
$|\varphi | $ varies very slowly as the universe evolves so
$\varphi$ is approximately constant during matter dominated era.
Variation of gravitational constant which is proportional to
$F_{1}$ in equation \ref{f1} is also small due to its
proportionality to $\frac{{1}}{{\omega }}$.

By using the ansatz, equation \ref{f1sol}, equation \ref{fi2}
gives the solution
\begin{equation}
F_{2}=F_{20}\left[ \frac{a}{a_{0}}\right] ^{\alpha } \label{f2sol}
\end{equation}%
\begin{equation}
\alpha =-\left( 3+\frac{1}{(1+\omega )}\right)  \label{alpha}
\end{equation}%
since $\omega >10^{4}$ for all practical purposes $\alpha =-3$.
For dust solution of complex JBD model, $\rho _{M}=\rho _{0}\left( \frac{a}{%
a_{o}}\right) ^{-3}$ and $p_{M}=0$, when equation \ref{f1sol} and
\ref{f2sol} are placed into equation \ref{density},
equations \ref{pressure}, \ref{fi1} are satisfied. Equation %
\ref{h2mat} is derived as,
\begin{eqnarray}
H^{2}=\frac{4\omega (1+\omega )^{2}}{(3\omega +4)(2\omega +3)}
\left[ m^{2}+\frac{2\rho _{0}}{\left\vert \varphi _{0}\right\vert
^{2}}\left(\frac{a}{a_{0}}\right) ^{\alpha }+F_{20}^{2} \left(
\frac{a}{a_{0}}\right) ^{2\alpha }\right]. \label{h2mat}
\end{eqnarray}

\bigskip Several interesting features emerge for the vacuum case, $\rho
_{M}=p_{M}=0$. Then $H^{2}$ contains only a constant term and a
$1/a^{6}$ term, respectively. Since $F_{2}$ is directly related to
the phase of the scalar field in equation \ref{f2}, the $1/a^{6}$
term is identified with the effect of the phase to the expansion
of the universe,
\begin{equation}
H^{2}(a)=\frac{4\omega (1+\omega )^{2}}{(3\omega +4)(2\omega +3)}%
(m^{2}+F_{2}^{2}(a)) .  \label{h2}
\end{equation}%
When we restrict the model to the $F_{2}=0$ case, $F=F_{1}$,
complex scalar field turns into a real scalar field and previously
studied JBD equations are obtained \cite{calik}. To test the
stability of our vacuum solution; we set
\begin{eqnarray}
a=a(1+\epsilon\eta)
\end{eqnarray}%
\begin{eqnarray}
 \varphi=\varphi(1+\epsilon\psi).
\end{eqnarray}
Inserting these variables into Eq.\ref{density}-\ref{fi2}, and
after neglecting the higher order terms in $\epsilon$, we obtain
four homogeneous differential equations for the three functions
$\eta, \psi_{R}$, and $\psi_I$. We set
\begin{eqnarray}
\eta=\eta_0e^{\beta t}
\end{eqnarray}%
\begin{eqnarray}
\psi_R=\psi_{R0}e^{\beta t}
\end{eqnarray}
\begin{eqnarray}
\psi_I=\psi_{I0}e^{\beta t}
\end{eqnarray}
and obtain four linear homogeneous equations for the three
unknowns $\eta_0, \psi_{R0}$, and $\psi_{I0}$. This homogeneous
system has a $4$x$3$ matrix of coefficients. The condition that a
nontrivial solution exists is that the rank of the matrix of
coefficients is at most two. All $3$x$3$ subdeterminants must be
zero to obtain a nontrivial solution for $\beta$.

We thus obtain four equations for the four $3$x$3$
subdeterminants, and arrange them in powers of $\beta$. One
equation is cubic in $\beta$ whereas the other three are
quadratic.
\begin{eqnarray}
A_{11}\beta^3+A_{12}\beta^2+A_{13}\beta+A_{14}=0
\end{eqnarray}
\begin{eqnarray}
A_{22}\beta^2+A_{23}\beta+A_{24}=0
\end{eqnarray}%
\begin{eqnarray}
A_{32}\beta^2+A_{33}\beta+A_{34}=0
\end{eqnarray}
\begin{eqnarray}
A_{42}\beta^2+A_{43}\beta+A_{44}=0
\end{eqnarray}

The determinant of the matrix of coefficients (det A) is nonzero,
so we can conclude that there is no solution for $\beta$ and it
means that the solution is stable.

 With the addition of the cosmological constant to cold dark matter(CDM) model, the
resulting $\Lambda$CDM model gives a better fit
\cite{komatsu,spergel}. Five-Year Wilkinson Microwave Anisotropy
Probe (WMAP) temperature and polarization observations \cite{wmap}
which include data from Baryon Acoustic Oscillations in the galaxy
and Type Ia supernova luminosity/time dilation measurements
\cite{kowalski} are used in the fitting process.

The constant term in equation \ref{h2mat} plays the role of the
cosmological constant and we will investigate the extra term from
the phenomenological point of view.

\section{Phenomenology}

\bigskip For standard cosmology in "matter dominated" era,
\begin{equation}
\frac{H^{2}}{H_{0}^{2}}=\Omega _{\Lambda }+\Omega
_{M}(\frac{a_{0}}{a})^{3} \label{pheno1}
\end{equation}%
where $\Omega _{\Lambda }$ is the fraction of vacuum energy and
the matter fraction, $\Omega _{M}$ is interpreted as $\Omega
_{M}=\Omega _{VM}+\Omega _{DM}$ where $\Omega _{VM}$ is the
fraction of visible matter and $\Omega _{DM}$ is the fraction of
dark matter. For the complex JBD model, Friedmann equation
becomes,
\begin{equation}
\left( \frac{H}{H_{0}}\right) ^{2}=\Omega _{\Lambda }+\Omega
_{M}\left( \frac{a_{0}}{a}\right) ^{3}+\Omega _{\Delta }\left(
\frac{a_{0}}{a}\right) ^{6}.  \label{pheno2}
\end{equation}%
We make the variable transformation and define
\begin{eqnarray}
u &=&\sqrt{\Omega _{\Lambda }}\left( \frac{a}{a_{0}}\right) ^{3}+\frac{%
\Omega _{M}}{2\sqrt{\Omega _{_{\Lambda }}}} \\
c &=&\sqrt{\mid \Omega _{\Delta }-\frac{\Omega _{M}^{2}}{4\Omega _{\Lambda }}%
\mid } \\
\varkappa &=&sgn(\Omega _{\Delta }-\frac{\Omega _{M}^{2}}{4\Omega _{\Lambda }%
})
\end{eqnarray}

so that Eq.\ref{pheno2} is put in differential form

\begin{equation}
3\sqrt{\Omega _{\Lambda }}H_{0}dt=\frac{du}{\sqrt{u^{2}+\varkappa
c^{2})}}. \label{diff}
\end{equation}

To test the viability of equation \ref{pheno2} predicted by
complex JBD model, we have to compare the standard fit to union
data \cite{kowalski} with a fit using equation \ref{pheno2} and
$H_{0}$=71 km/sec/Mpc. With the constraint $\Omega _{\Lambda
}+\Omega _{M}=1$, the standard model fit has one free parameter
which can be chosen as $\Omega _{\Lambda }$.

\bigskip The latest Type Ia supernova (SNIa) data sets in Table C2 are taken
by the Supernova Cosmology Project Group \cite{kowalski}. $414$
SNIa supernova magnitude-redshift observations are compiled with
"Union" and after selection cuts, it reduces to $307$ Sne.
Luminosity distance, in Friedmann-Robertson-Walker Cosmology, is
defined in \cite{perlmutter}
\begin{eqnarray}
D_{L}=\frac{c(1+z)}{H_{o}}\int dz^{\prime }[\sum \Omega
_{i}(1+z^{\prime })^{3(1+w_{i})}\nonumber -\kappa _{o}(1+z^{\prime
})^{2}]^\frac{-1}{2} \label{dist}
\end{eqnarray}

where $\kappa _{o}=\sum \Omega _{i}-1$.

Distance modulus can be written in terms of luminosity distance%
\begin{equation}
\mu =m-M=5\log (\frac{D_{L}}{Mpc})+25,  \label{mag}
\end{equation}%
where m is the apparent magnitude and M is the absolute magnitude.
Three different cases for $\varkappa$ are analyzed. Three fits we
would like to present are;

\bigskip $1%
{{}^\circ}%
$) $\varkappa =-1$, $\Omega _{\Delta }$=0 standard cosmology fit with $%
\Omega _{\Lambda }+\Omega _{M}=1$ gives,
\begin{equation}
\left( \frac{H}{H_{0}}\right) ^{2}=0.745+0.255\left(
\frac{a_{0}}{a}\right) ^{3}  \label{fit1}
\end{equation}%
with $\chi ^{2}/d.o.f.=1.45$. The scale factor and lifetime of
universe are
related by,%
\begin{equation}
\left( \frac{a}{a_{0}}\right) ^{3}=\frac{\Omega _{M}}{2\Omega _{\Lambda }}%
(\cosh (3H_{0}\sqrt{\Omega _{\Lambda }}t)-1)
\end{equation}

\bigskip $2%
{{}^\circ}%
$) $\varkappa =+1$ fit to complex JBD model with $\Omega _{M}=0$
and $\Omega _{\Lambda }+\Omega _{\Delta }=1$ gives,
\begin{equation}
\left( \frac{H}{H_{0}}\right) ^{2}=0.938+0.062\left(
\frac{a_{0}}{a}\right) ^{6}  \label{fit2}
\end{equation}%
with $\chi ^{2}/d.o.f.=1.43$. The scale factor and lifetime of
universe are
related by,%
\begin{equation}
\left( \frac{a}{a_{0}}\right) ^{3}=\sqrt{\frac{\Omega _{\Delta
}}{\Omega _{\Lambda }}}\sinh (3H_{0}\sqrt{\Omega _{\Lambda }}t).
\end{equation}

\bigskip $3%
{{}^\circ}%
$) $\varkappa =0$ , fit to complex JBD model with matter, $\Omega
_{\Lambda }+\Omega _{M}+\Omega _{\Delta }=1$ and $\Omega _{\Delta
}=\frac{\Omega _{M}^{2}}{4\Omega _{\Lambda }}$ so that \ equation
\ref{pheno2} becomes
\begin{equation}
\frac{H}{H_{0}}=\sqrt{\Omega _{\Lambda }}+\sqrt{\Omega _{\Delta }}(\frac{%
a_{0}}{a})^{3}  \label{pheno4}
\end{equation}%
gives,
\begin{equation}
\left( \frac{H}{H_{0}}\right) ^{2}=0.790+0.200\left(
\frac{a_{0}}{a}\right) ^{3}+0.010\left( \frac{a_{0}}{a}\right)
^{6}  \label{fit3}
\end{equation}%
with $\chi ^{2}/d.o.f.=1.43$. This condition makes the r.h.s. of
Friedmann equation perfect square and gives the best fit to
supernovae union data. The scale factor and lifetime of universe
are related by,

\begin{equation}
\left( \frac{a}{a_{0}}\right) ^{3}=\frac{\Omega _{M}}{2\Omega
_{\Lambda }} (\exp (3H_{0}\sqrt{\Omega _{\Lambda }}t)-1)
\end{equation}

\begin{figure}[tbp]
\centering \includegraphics[width=9cm,height=7cm]{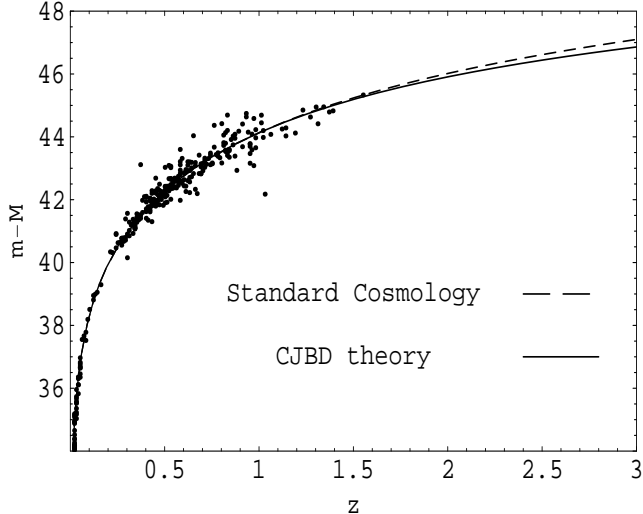}
\caption{Magnitude vs. Redshift Graph} \label{fig1.eps}
\end{figure}
\bigskip The fit for equation \ref{fit2} is interesting but unacceptable
since the model estimated lifetime is approximately 10 Gyrs, small
compared with the observations \cite{Suyu}. In fig.1, standard
cosmology fit using Eq.\eqref{fit1} and complex JBD fit
Eq.\eqref{fit3} to Union data sets \cite{kowalski} are shown. We
conclude that the extra term, from the change of the JBD field's
phase,which must be small, indicates a slightly better fit. The
difference between two models will be seen for high redshift
observations. Similar to the addition of cosmological constant in
$\Lambda $CDM, cosmological reason of obtaining the phase term in
JBD theory can be investigated.

\section{Conclusion and Acknowledgements}

\bigskip Complex JBD model which in addition to the constant term makes
a contribution of $\frac{1}{a^{6}}$ term to the Friedmann Equation
fits the Supernovae data accurately. It is clear from \ref{h2},
which is valid for  $\rho _{M}=p_{M}=0$, that this term is a
natural component of dark energy. Actually we conclude that the
complex JBD model explains the evolution of the universe with a
slightly better fit. However; the lifetime of the universe was
found approximately 12 Gyrs, smaller than the experimental age of
the universe, 13,8 Gyrs \cite{Suyu}. Moreover; the complex scalar
field is responsible in large scale for the expansion behavior of
universe and its phase behaves as a phenomenological extreme
density term. The more the density due to complex phase becomes, a
smaller model-based age is determined. In standard cosmology, the
$\frac{1}{a^{6}}$ term would be obtained if the kinetic term of a
scalar field dominates the energy-momentum tensor. In our model,
this is also how it mathematically arises. However, the physical
interpretation is that it is, as a typical feature of CJBD theory,
a component of dark energy and its presence may be determined by
more accurate measurements of the Supernova data near $z\approx3$.
This is far from the radiation dominated age, $z\approx1100$. This
research is in part supported by The Turkish Academy of Sciences,
TUBA.

\end{document}